\documentstyle[12pt,epsfig]{article}
\evensidemargin=\oddsidemargin
\def\thefootnote{\fnsymbol{footnote}}

\def\mathrm#1{\mbox{\rm #1}}
\def\bold#1{\setbox0=\hbox{$#1$}%
     \kern-.025em\copy0\kern-\wd0
     \kern.05em\copy0\kern-\wd0
     \kern-.025em\raise.0433em\box0 }
\def\21{$SU(2) \otimes U(1)$}
\newcommand{\matriz}{\left[\begin{array}} 
\newcommand{\finmatriz}{\end{array}\right]} 
\def\frad#1#2{\frac{\displaystyle{#1}}{\displaystyle{#2}}}
\def\frap#1#2{{\hbox{$\frac{#1}{#2}$}}}
\def\half{{\textstyle{1 \over 2}}}
\def\etal{\hbox{\it et al., }}
\def\eq#1{{eq. (\ref{#1})}}
\def\VEV#1{\left\langle #1\right\rangle}
\def\lsim{\raise0.3ex\hbox{$\;<$\kern-0.75em\raise-1.1ex\hbox{$\sim\;$}}}
\def\gsim{\raise0.3ex\hbox{$\;>$\kern-0.75em\raise-1.1ex\hbox{$\sim\;$}}}
\def\mpl#1#2#3{          {\it Mod. Phys. Lett. }{\bf #1} (19#2) #3}
\def\np#1#2#3{           {\it Nucl. Phys. }{\bf #1} (19#2) #3}

\def\pl#1#2#3{           {\it Phys. Lett. }{\bf #1} (19#2) #3}
\def\ppnp#1#2#3{           {\it Prog. Part. Nucl. Phys. }{\bf #1} (19#2) #3}
\def\pr#1#2#3{           {\it Phys. Rev. }{\bf #1} (19#2) #3}
\def\prep#1#2#3{         {\it Phys. Rep. }{\bf #1} (19#2) #3}
\def\prl#1#2#3{          {\it Phys. Rev. Lett. }{\bf #1} (19#2) #3}

\begin{document}
\begin{titlepage}
\pagestyle{empty}
\rightline{FTUV/97-54}
\rightline{IFIC/97-55}
\rightline{hep-ph/9803362}
\rightline{\today}
%\hfill Submitted to Nucl. Phys. B
\vskip 1.0cm
\begin{center}
{\bf SEESAW MAJORON MODEL OF NEUTRINO MASS AND NOVEL SIGNALS IN HIGGS
BOSON PRODUCTION AT LEP }
\vskip 1.cm
{\large Marco A. D\'\i az, 
M. A. Garc{\'\i}a-Jare\~no,} \\
{\large Diego A. Restrepo and 
 Jos\'e W. F. Valle \footnote{E-mail: mad@flamenco.ific.uv.es, 
miguel@flamenco.ific.uv.es, restrepo@flamenco.ific.uv.es, and \\
valle@flamenco.ific.uv.es}}\\
\hspace{3cm}\\
{\sl
Departamento de F\'\i sica Te\'orica, IFIC-CSIC, Universidad de Valencia}\\ 
{\sl Burjassot, Valencia 46100, Spain}
\vskip 1.cm
\baselineskip 14pt
\begin{quotation}
We perform a careful study of the neutral scalar sector of a model
which includes a singlet, a doublet, and a triplet scalar field under
$SU(2)$. This model is motivated by neutrino  physics, since it is simply
the most general version of the seesaw model of neutrino mass
generation through spontaneous violation of lepton number. The neutral
Higgs sector contains three CP-even and one massive CP-odd Higgs boson
$A$, in addition to the massless CP-odd majoron $J$.  The weakly
interacting majoron remains massless if the breaking of lepton number
symmetry is purely spontaneous. We show that the massive CP-odd Higgs
boson may invisibly decay to three majorons, as well as to a CP-even
Higgs $H$ boson plus a majoron. We consider the associated Higgs
production $e^+e^- \to Z \to H A$ followed by invisible decays $A \to
JJJ$ and $H \to JJ$ and derive the corresponding limits on masses and
coupling that follow from LEP I precision measurements of the
invisible Z width. We also study a novel $b \bar{b}
b\bar{b}p\!\!\!/_T$ signal predicted by the model, analyse the
background and perform a Monte-Carlo simulation of the signal in order
to illustrate the limits on Higgs boson mass, couplings and branching
ratios that follow from that.
\end{quotation}
\end{center}
\end{titlepage}
\vfill
\noindent
\def\thefootnote{\arabic{footnote}}
\setcounter{footnote}{0}
\setcounter{page}{1}
\pagestyle{plain}
\baselineskip 18pt

\section{Introduction}

One of main puzzles in particle physics is the origin of mass in
general as well as the problem of neutrino mass in particular. In the
Standard Model (SM) the spontaneous breaking of the gauge symmetry
through the expectation value of a scalar \21 doublet is the origin of
the masses of the fermions as well as those of the gauge bosons. The
main implication for this scenario is the existence of the Higgs boson
\cite{HIGGS}, not yet found \cite{RichardVal95,lep2}.  Many of the
extensions of the Standard Model Higgs sector postulated in order to
generate mass for neutrinos are characterised by the spontaneous
violation of a global $U(1)$ lepton number symmetry by an \21 singlet
vacuum expectation value $\VEV{\sigma}$ \cite{fae}. These models
contain a massless Goldstone boson, called majoron ($J$), which
interacts very weakly with normal matter \cite{CMP}. Although the
interactions of the majoron with quarks, leptons, and gauge bosons is
naturally very weak, as required also by astrophysics \cite{KIM}, it
can have a relatively strong interaction with the Higgs boson
\cite{HJJ,JR}. It has been noted that the main Higgs boson decay
channel is likely to be {\sl invisible}, e.g.
\begin{equation} 
H \to J J \; ,
\end{equation} 
where $J$ denotes the majoron field. This feature also appears in
variants of the minimal supersymmetric model in which $R$ parity is
broken spontaneously \cite{MASIpot3}. The phenomenological
implications of the invisible CP-even Higgs boson decays for
various collider experiments have been extensively discussed
\cite{valle:1,dproy,valle:2,valle:asso,eboli,l3inv}.

In the seesaw model \cite{GRS,LR} one adds an \21 isosinglet
right-handed neutrino associated with each generation of isodoublet
neutrinos.  In addition to the standard lepton number conserving {\sl
isodoublet} mass term analogous to those responsible for the charged
fermion masses, there is also a Majorana mass term for the right
handed neutrinos and left-handed neutrinos. The neutrino mass matrix
takes the form
\begin{equation}
\left( 
\begin{array}{cc}
\nu  & \nu^c
\end{array}
\right)^T 
\sigma_2
\left( 
\begin{array}{cc}
M_L & D \\ 
D^T & M_R
\end{array}
\right)  \left( 
\begin{array}{c}
\nu  \\ 
\nu ^c
\end{array}
\right)  
\end{equation}  
where $\sigma_2$ is the charge conjugation matrix and the entries obey
the hierarchy $M_R >> D >> M_L$ \cite{2227}. In a model where
neutrinos acquire mass only from the spontaneous violation of lepton
number the entries $M_L$ and $M_R$ arise from vacuum expectation
values (VEV) of \21 triplet and singlet Higgs scalars $\Delta$ and
$\sigma$ \cite{774}, while the Dirac mass term $D$ follows from the
Standard Model doublet VEV.  In this model the light neutrino masses
arise from diagonalizing out the heavy right-handed neutrinos and has
a contribution from the small triplet VEV.

In this paper we show that, for a wide choice of parameters, the
complete version of the seesaw majoron model of neutrino mass
containing \21 doublet, singlet as well as triplet Higgs multiplets
(called {\bold 123} model in ref.\cite{774}) implies that the massive
pseudoscalar Higgs boson can also decay invisibly, either directly as
\begin{equation} 
\label{new1}
A \to 3 J \; ,
\end{equation} 
or indirectly as
\begin{equation} 
\label{new2}
A \to H J \; \mbox{with} \: \: \:\:  H \to J J \; 
\end{equation} 
when $m_A > m_H$. This feature has not been noted in any of the
discussions given so far \cite{valle:1,dproy,valle:asso,eboli}, as it
is absent in a number of models, for example the all the models
discussed in \cite{HJJ}.

Massive invisibly decaying CP-odd Higgs bosons may occur in the
minimal supersymmetric standard model, where the decay involves a
heavy fermion pair, $A \to \chi^0 \chi^0$, with $\chi^0$ stable due to
R-parity conservation. Similarly, it can also occur in the
supersymmetric model with spontaneously broken R-parity considered in
ref. \cite{MASIpot3}. In the latter case one could have, e.g.  $A \to
\chi^0 \chi^0$ or $A \to \nu \chi^0$, with $\chi^0 \to \nu J$, where
$J$ denotes the majoron. However, all these decays require a
kinematical condition $m_A > 2 m_{\chi^0}$ or $m_A > m_{\chi^0}$ which
is avoided here due to the majoron being massless.

We carefully study the scalar potential of the model and derive from
it the relevant CP-even as well as CP-odd Higgs boson mass
matrices. In the next section, we discuss the parameterisation of
Higgs bosons couplings relevant for their production at LEP, both for
the $Z H$ as well as $A H$ production channels.  Next we use these
theoretical results in order to derive restrictions on the relevant
Higgs boson parameters from the precision measurements of the
invisible width of the Z boson at LEP I. The associated production
$e^+e^- \to H A$ with $A \to H J$ and $H \to b \bar{b}$ also leads to
a novel $b \bar{b} b\bar{b}p\!\!\!/_T$ signal topology that could be
detectable at LEP II. We have performed a detailed analysis of the
background and carried out a Monte-Carlo simulation of the signal in
order to illustrate the limits on Higgs boson mass, couplings and
branching ratios that follow from four-jet + missing momentum signal
topology.

\section{The Scalar Potential}

The model we consider here is the one proposed in ref. \cite{774} as a
generalisation of the triplet \cite{GelRon} and singlet \cite{CMP}
majoron models.  The Higgs sector of the model contains the usual
$SU(2)$ Higgs complex doublet $\phi$ of the SM, with lepton number
$L=0$, and an $SU(2)$ complex triplet $\Delta$, with lepton number
$L=-2$,
\begin{equation}
\Delta=\matriz{cc} \Delta^0&\Delta^+/\sqrt2\\
\Delta^+/\sqrt2&\Delta^{++}\finmatriz,
\qquad\qquad\phi=\matriz{c}\phi^0\\
\phi^-\finmatriz,
\label{higgs}
\end{equation}
where we have used the $2\times 2$ matrix notation for the Higgs
triplet given in ref. \cite{GGN}. The Higgs sector is completed with a
complex \21 singlet scalar, denoted $\sigma$, carrying lepton number
$L=2$.

The full scalar potential is given by \cite{joshipura}
\def\tr{{\rm tr}\,}
\begin{eqnarray}
V(\phi,\Delta,\sigma)&=&\mu_2^2\phi^\dagger\phi+\mu_3^2\,
\tr(\Delta^\dagger\Delta)+
\lambda_1(\phi^\dagger\phi)^2+\lambda_2[\tr(\Delta^\dagger\Delta)]^2
\nonumber\\
&&+\lambda_3\phi^\dagger\phi\,\tr(\Delta^\dagger\Delta)+
\lambda_4\,\tr(\Delta^\dagger\Delta\Delta^\dagger\Delta)+
\lambda_5(\phi^\dagger\Delta^\dagger\Delta\phi)\nonumber\\
&&+\mu_1^2\sigma^\dagger\sigma+\beta_1(\sigma^\dagger\sigma)^2+
\beta_2(\phi^\dagger\phi)(\sigma^\dagger\sigma) \nonumber\\
&&+\beta_3\,\tr(\Delta^\dagger\Delta)\sigma^\dagger\sigma-
\kappa(\phi^T\Delta\phi\sigma+{\rm h.c.})
\label{eq:GenPot}
\end{eqnarray}
where $\mu_i$, $i=1,2,3$, are mass parameters, and $\lambda_i$,
$i=1,...5$, $\beta_i$, $i=1,2,3$, and $\kappa$ are dimensional-less
couplings. The first two lines in eq.~(\ref{eq:GenPot}) correspond to
the Gelmini--Roncadelli model \cite{GelRon}, and the last two lines
are new terms involving the scalar $\sigma$. The term in $\kappa$ has
been introduced in ref. \cite{774} and plays an important role in our
present discussion.

The singlet $\sigma$, as well as the neutral components of the fields 
$\phi$ and $\Delta$, acquire vacuum expectation values $v_1$, $v_2$, and 
$v_3$ respectively. According to this, we shift the fields in the 
following way
\begin{eqnarray}
\sigma&=&\frad{v_1}{\sqrt2}+\frad{R_1+iI_1}{\sqrt{2}}\nonumber\\
\phi^0&=&\frad{v_2}{\sqrt{2}}+\frad{R_2+iI_2}{\sqrt{2}}
\label{shift}\\
\Delta^0&=&\frad{v_3}{\sqrt2}+\frad{R_3+iI_3}{\sqrt{2}}\nonumber
\end{eqnarray}
We assume that the three vacuum expectation values are real.
The scalar potential contains the following linear terms
\begin{equation}
V_{linear}=t_1R_1+t_2R_2+t_3R_3\,,
\label{eq:Vlinear}
\end{equation}
where
\begin{eqnarray}
t_1&=&v_3(\mu_1^2+\lambda_2v_3^2+\half\lambda_3v_2^2+\lambda_4v_3^2+
\half\lambda_5v_2^2+\half\beta_3v_1^2)-\half\kappa v_1v_2^2\nonumber\\
t_2&=&v_1(\mu_2^2+\beta_1v_1^2+\half\beta_2v_2^2+\half\beta_3v_3^2)-
\half\kappa v_2^2v_3
\label{eq:tadpoles}\\
t_3&=&v_2(\mu_3^2+\lambda_1v_2^2+\half\lambda_3v_3^2+\half\lambda_5
v_3^2+\half\beta_2v_1^2-\kappa v_1v_3)
\nonumber
\end{eqnarray}
The conditions for a extreme of the potential are $t_i=0$, $i=1,2,3$.
Therefore, the $t_i=0$ vanish at the minima of the potential. We will
verify explicitly below that, for many choices of its parameters, the
potential has indeed minima for nonzero values of $v_1$, $v_2$ and $ v_3$.

\section{Neutral Higgs Mass Matrices}

Taking into account the fact that this model contains one
doubly-charged and and one singly-charged scalar boson, in addition to
the two charged unphysical \21 Goldstone modes (longitudinal W), it
follows that the neutral Higgs sector of this model is composed by six
real fields. Due to CP invariance they split into two unmixed sectors
of three CP--even and three CP--odd fields.  Their mass matrices are
contained in the quadratic scalar potential which includes:
\begin{equation}
V_{quadratic}=\half\Big[R_1,R_2,R_3\Big]
{\bold M^2_R}\left[\matrix{
R_1 \cr R_2 \cr R_3
}\right]
+\half\Big[I_1,I_2,I_3\Big]
{\bold M^2_I}\left[\matrix{
I_1 \cr I_2 \cr I_3
}\right]+...
\label{eq:NeutScalLag}
\end{equation}
The CP--even Higgs mass matrix, which is in agreement with ref.
\cite{joshipura}, is given by
\begin{equation}
\label{MR}
{\bold M_R^2}=\left[
\begin{array}{ccc}
2\beta_1v_1^2+\half\kappa v_2^2\frad{v_3}{v_1}+\frad{t_1}{v_1}&
\beta_2v_1v_2-\kappa v_2v_3 
&\beta_{3}v_1v_3-\half\kappa v_2^2\\
\beta_2v_1v_2-\kappa v_2v_3&2\lambda_1v_2^2+\frad{t_2}{v_2}
&(\lambda_3+\lambda_5)v_2v_3-\kappa v_1v_2\\
\beta_{3}v_1v_3-\half\kappa v_2^2&(\lambda_3+\lambda_5)v_2v_3-\kappa v_1v_2
&2(\lambda_2+\lambda_4)v_3^2+\half\kappa v_2^2\frad{v_1}{v_3}+
\frad{t_3}{v_3}\\
\end{array}  \right]
\end{equation}
where it is safe to take $t_i=0$, $i=1,2,3$, unless $v_1=0$ or $v_3=0$
in which case the expression of the corresponding extremization
condition (``tadpole equation'') in eq.~(\ref{eq:tadpoles}) must be
replaced in the mass matrix ${\bf M}_R^2$.  The physical CP--even mass
eigenstates $H_i$, $i=1,2,3$, are related to the corresponding weak
eigenstates $R_i$ as
\begin{equation}
\label{me}
\matriz{c}
H_1\\H_2\\H_3
\finmatriz=O_R\;\matriz{c}
R_1\\R_2\\R_3
\finmatriz.
\end{equation}
where the $3\times3$ matrix $O_R$ is the matrix which diagonalizes the 
CP-even mass matrix such that
\begin{equation}
O_R{\bold M_R^2}O_R^T={\rm diag}(m_{H_1}^2,m_{H_2}^2,m_{H_3}^2)
\label{eq:diagMR}
\end{equation}
and where by definition we take $m_{H_1}\le m_{H_2}\le m_{H_3}$. 

The CP--odd Higgs mass matrix ${\bold M_I^2}$ is given by \cite{774}
\begin{equation}
{\bold M_I^2}=\matriz{ccc}
\half\kappa v_2^2\frad{v_3}{v_1}+\frad{t_1}{v_1} & \kappa v_2v_3 & 
\half\kappa v_2^2\\
\kappa v_2v_3 & 2\kappa v_1v_3+\frad{t_2}{v_2} & \kappa v_1v_2 \\
\half\kappa v_2^2 & \kappa v_1v_2 & \half\kappa 
v_2^2\frad{v_1}{v_3}+\frad{t_3}{v_3}
\finmatriz .
\label{eq:MassMatrixI}
\end{equation}
If $v_1\neq 0$ and $v_3\neq 0$ we can safely set the tadpoles equal to zero
in eq.~(\ref{eq:MassMatrixI}), in whose case we can see that the matrix
${\bold M_I^2}$ has two zero eigenvalues. One of them is the unphysical 
Goldstone boson and the other is the physical Majoron.
The physical CP--odd mass eigenstates $A_i$, $i=1,2,3$, are related to the 
corresponding weak eigenstates $I_i$ as
\begin{equation}
\label{modd}
\matriz{c}
A_1\\A_2\\A_3
\finmatriz
\equiv
\matriz{c}
J\\G\\A
\finmatriz
=O_I\;\matriz{c}
I_1\\I_2\\I_3
\finmatriz.
\end{equation}
where the $3\times3$ matrix $O_I$ is the matrix which diagonalizes the 
CP-odd mass matrix such that
\begin{equation}
O_I{\bold M_I^2}O_I^T={\rm diag}(0,0,m_A^2)
\label{eq:diagMI}
\end{equation}
and the CP--odd Higgs mass is given by
\begin{equation}
m_A^2=\half\kappa\frad{v_2^2v_1^2+v_2^2v_3^2+4v_3^2v_1^2}{v_3v_1}
\label{MA}
\end{equation}
Note that $m_A \to 0$ as $\kappa \to 0$. The diagonalization matrix
$O_I$ can be found analytically
\begin{equation}
\label{med}
O_I=\matriz{ccc}
cv_1V^2           & -2cv_2v_3^2 & -cv_2^2v_3        \\
0                 & \frad{v_2}V & -2\frad{v_3}V     \\
b\frad{v_2}{2v_1} & b           & b\frad{v_2}{2v_3}
\finmatriz,
\end{equation}
where $V$, $c$, and $b$ are the normalisation constants for the 
eigenvectors $G$, $J$, $A$ respectively. They are given by
\begin{eqnarray}
V^2    &=& v_2^2+4v_3^2 \nonumber\\
c^{-2} &=& v_1^2V^4+4v_2^2v_3^4+v_2^4v_3^2 \label{eq:normalization}\\
b^2    &=& \frad{4v_1^2v_3^2}{v_2^2v_1^2+v_2^2v_3^2+4v_3^2v_1^2}
\nonumber
\end{eqnarray}

We now briefly discuss three special cases, motivated by the cases when
tadpoles cannot be trivially set to zero in the scalar mass matrices
${\bold M_R^2}$ and ${\bold M_I^2}$.
\begin{itemize}
\item $v_1=0$, $v_3=0$.  \\
In this case there is no breaking of lepton number, as in the Standard
Model and, as a result there is no massless Majoron.  The unphysical
Goldstone boson is purely doublet. One of the CP--even Higgs bosons is
also pure doublet with a mass $m_H^2=2\lambda_1v_2^2$. The remaining
two CP--even Higgs bosons are massive and are a mixture of singlet and
triplet. There are also two massive CP--odd Higgs bosons and they are
degenerate with the CP--even Higgs bosons.  
\item $v_1\neq 0$, $v_3=0$.  
In this case the third minimization equation forces to have
$\kappa=0$.  There is a Majoron with $m_J=0$ which is purely singlet,
as in the simplest {\bold 12} model considered in ref. \cite{774}, and
the unphysical Goldstone boson is purely doublet. The real and
imaginary parts of the neutral component of the triplet field are
degenerate and form a complex field with mass $m_{\Delta^0}^2 =
\mu_3^2+\half(\lambda_3+\lambda_5)v_2^2+\half\beta_3v_1^2$.  There are
two additional massive CP--even Higgs bosons, mixture of singlet and
doublet.
\item $v_1=0$, $v_3\neq 0$.\\
In this case the first tadpole equation forces to have $\kappa=0$.
There is a Majoron with $m_J=0$ which has zero component along the
singlet, and is therefore experimentally ruled out by the LEP data.
Here the situation is analogous to the simplest {\bold 23} model of
ref. \cite{774} and it is for this reason that the presence of the
singlet field $\sigma$ with non-zero VEV is mandatory.
\end{itemize}
Thus we conclude that the situation of interest for us is the general
one in which all three VEVs $v_1$, $v_2$ and $ v_3$ take on nonzero
values. In our numerical calculations reported in section V we must
take into account an important astrophysical constraint on these VEVs
that follows from stellar cooling by majoron emission which severely
restricts the majoron electron coupling to be less than about
$10^{-12}$ or so. This is discussed in detail in section V.

\section{Higgs Boson Production and Decays}

In this section we derive the couplings relevant for Higgs boson
production at $e^+e^-$ colliders and for their invisible decays. The
two production mechanisms we consider are the emission of a CP--even
Higgs $H$ by a $Z$--boson, and the associated production consisting of
a $Z$--boson decaying into a CP--even Higgs $H$ and a CP--odd Higgs
$A$.  In order to derive the couplings $ZZH$ and $ZHA$, we need the
kinetic part of the scalar Lagrangian contained in
\begin{equation}
{\cal L}_{\rm scalar}=({\cal D}_\mu\phi)^\dagger{\cal D}^\mu\phi
+{\rm tr}\,[({\cal D}_\mu\Delta)^\dagger{\cal D}^\mu\Delta]+
\partial_\mu\sigma^\dagger\partial^\mu\sigma-V(\phi,\Delta,\sigma),
\label{der}
\end{equation}
where the covariant derivative is defined by
\begin{equation}
{\cal D}=\partial^\mu+ig{\bf T}\cdot{\bf W}^\mu+\frad i2g'YV^\mu
\end{equation}
and $g$ and $g'$ are the gauge couplings corresponding to the $SU(2)$
and $U(1)$ groups respectively.
On the scalars fields, the generators act as follows
\begin{eqnarray}
{\bf T}\phi&=\frad12{\vec\tau}\phi,\qquad
{\bf T}\Delta&=-\frad12{\vec\tau}\Delta-\frad12\Delta{\vec\tau}^*\nonumber\\
Y\phi&=-1\phi,\qquad Y\Delta&=2\Delta\,,
\label{WZ}
\end{eqnarray}
and with these definitions we have $T_3\phi^0=\frad12\phi^0$ and 
$T_3\Delta^0=-1\Delta^0$.

The Higgs singlet does not contribute to the gauge boson masses, therefore,
they depend only on $v_2$ and $v_3$ and are given by \cite{2227}
\begin{equation}
m_Z^2=\frad{g^2}{4\cos^2\theta_W}(v_2^2+4v_3^2),\qquad
m_W^2=\frad{g^2}{4}(v_2^2+2v_3^2)
\label{mw}
\end{equation}
and from the measurement of
the $\rho$ parameter one has a restriction on $v_3$, namely \cite{pichval97}
\begin{equation}
\rho=1+\frac{2v_3^2}{v_2^2+2v_3^2}=1.001\pm 0.002\,.
\label{rho}
\end{equation}
which implies in practice that $v_3\leq 8$ GeV.

We mow calculate the relevant couplings for the production of Higgs
bosons at $e^+e^-$ colliders.
Using eq.~(\ref{der}), we determined the $HAZ$ couplings to be
\begin{equation}
{\cal L}_{HAZ}=\frad g{2c_w}Z^\mu\left[\,R_2
\stackrel{\longleftrightarrow}{\partial^\mu}\,I_2-2\,R_3
\stackrel{\longleftrightarrow}{\partial^\mu}\,I_3\right]
=\frad g{2c_w}Z^\mu\left[
{O^I_{32}}O^R_{a2}-2{O^I_{33}}O^R_{a3}\right]\,H_a
\stackrel{\longleftrightarrow}{\partial^\mu}\,A.
\label{ZHAa}
\end{equation}
where $c_W\equiv\cos\theta_W$ and $H_a$ is any of the three CP--even
neutral Higgs bosons. The quantity defined by
\begin{equation}
\epsilon_A={O^I_{32}}O^R_{12}-2{O^I_{33}}O^R_{13}
\end{equation}
essentially represents the strength of the coupling $H_1AZ$ [see the
second squared parenthesis in eq.~(\ref{ZHAa})]. 

\begin{figure}
\centerline{\protect\hbox{
\psfig{file=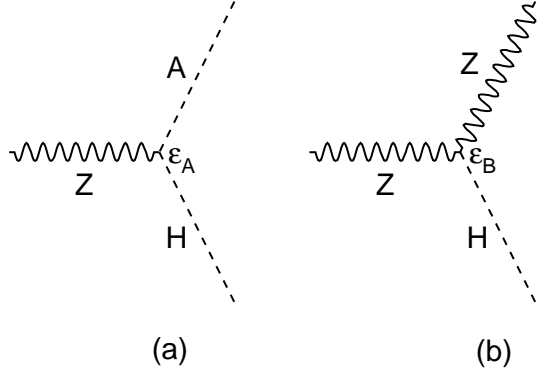,height=10cm}}}
\caption{Feynman rules relevant to Bjorken and Associated Higgs 
Production.}
\label{bjorka}
\end{figure}

{}From eq.~(\ref{der}) we find that the $HZZ$ coupling is 
\begin{equation}
{\cal L}_{HZZ}={g\over{4c_W^2}}m_ZZ^\mu Z_\mu
\left[\frac{v_2}VO^R_{a2}+\frac{4v_3}VO^R_{a3}\right]H_a,
\end{equation}
and again we define the following quantity
\begin{equation}
\epsilon_B=\frac{v_2}VO^R_{12}+\frac{4v_3}VO^R_{13}
\label{eps_B}
\end{equation}
as a measure of the strength of the $H_1ZZ$ coupling. Therefore, the 
Bjorken production mechanism (Fig.~\ref{bjorka}a) and the associated
production mechanism (Fig.~\ref{bjorka}b) are determined by the 
parameters $\epsilon_A$ and $\epsilon_B$ respectively.

We now turn to the couplings relevant for the invisible decay of the
Higgs bosons $H_1$ and $A$. We need to find first the trilinear 
couplings $HJJ$, $HAJ$, and $HAA$, thus we start with the cubic part 
of the scalar potential which involve one (unrotated) CP--even Higgs 
field $R_i$ and two (unrotated) CP-odd Higgs fields $I_j$. This part 
of the potential is given by
\begin{eqnarray}
V_{RII}&=&\lambda_1v_2I_2^2R_2+(\lambda_2+\lambda_4)v_3I_3^2R_3+
\half(\lambda_3+\lambda_5)(v_2I_3^2R_2+v_3I_2^2R_3)
\nonumber\\
&&+\beta_1v_1I_1^2R_1+\half\beta_2(v_2I_1^2R_2+v_1I_2^2R_1)
+\half\beta_3(v_1I_3^2R_1+v_3I_1^2R_3)
\label{jjh}\\
&&+\kappa[v_1(\half I_2^2R_3+I_2I_3R_2)+v_3(\half I_2^2R_1+I_1I_2R_2)
+v_2(I_1I_2R_3+I_3I_1R_2+I_2I_3R_1)]
\nonumber
\end{eqnarray}
Making the substitution $I_iI_j\to O^I_{1i}O^I_{1j}J^2$ in
eq.~(\ref{jjh}), we find the coupling $H_aJJ$ in terms of the
mass $m_{H_a}$ and the rotation matrices $O^R$ and $O^I$
\begin{eqnarray} 
V_{HJJ}&=&
\frad12\left[\frad{{O^I_{12}}^2}{v_2}(M_R^2)_{1a}+
\frad{{O^I_{13}}^2}{v_3}(M_R^2)_{2a}+\frad{{O^I_{11}}^2}{v_1}
(M_R^2)_{3a}\right]R_aJ^2\nonumber\\
&=&\frad12\left[\frad{{O^I_{12}}^2}{v_2}{(O^R)}^T_{2a}+
\frad{{O^I_{13}}^2}{v_3}{(O^R)}^T_{3a}
+\frad{{O^I_{11}}^2}{v_1}{(O^R)}^T_{1a}\right]m_{Ha}^2H_aJ^2.
\end{eqnarray}
where eq.~(\ref{eq:diagMR}) has been used.

In a similar way, the terms in the Lagrangian relevant for the 
vertices $HAJ$ and $HAA$ can be found from eq.~(\ref{jjh}) making the
following substitutions
\begin{eqnarray}
V_{HAJ}&=&V_{RII}(I_iI_j\to [O^I_{3i}O^I_{1j}+
O^I_{1i}O^I_{3j}]AJ)\,,
\nonumber\\
V_{HAA}&=&V_{RII}(I_iI_j\to  O^I_{3i}O^I_{3j}A^2)\,.
\label{eq:HAJandHAA}
\end{eqnarray}
These terms are not explicitly displayed.

Finally we turn to the quartic coupling responsible for the invisible 
decay $A\to 3J$. The relevant piece of the quartic scalar potential is
\begin{equation}
V_{I^4}=\frap14[\lambda_1I_2^4
+(\lambda_2+\lambda_4)I_3^4+(\lambda_3+\lambda_5)I_2^2I_3^2
+\beta_1I_1^4+\beta_2I_2^2I_1^2+\beta_3I_3^2I_1^2
-2\kappa I_1I_2^2I_3]
\end{equation}
and after making the following substitution 
\begin{equation}
I_iI_jI_k^2\longrightarrow(O^I_{3i}O^I_{1j}O^{I2}_{1k}+
O^I_{1i}O^I_{3j}O^{I2}_{1k}+2O^I_{1i}O^I_{1j}O^I_{3k}O^I_{1k})AJ^3
\label{eq:subsI4}
\end{equation}
we find the term $AJ^3$ in the scalar potential:
\begin{eqnarray}
V_{AJ^3}&=&\Big[
\lambda_1{O_{12}^I}^3O_{32}^I
+(\lambda_2+\lambda_4){O_{13}^I}^3O_{33}^I+\half
(\lambda_3+\lambda_5)O_{12}^IO_{13}^I(O_{32}^IO_{13}^I+O_{33}^IO_{12}^I)
\nonumber\\ &&
\beta_1{O_{11}^I}^3O_{31}^I+
+\half\beta_2O_{11}^IO_{12}^I(O_{31}^IO_{12}^I+O_{32}^IO_{11}^I)
+\half\beta_3O_{11}^IO_{13}^I(O_{31}^IO_{13}^I+O_{33}^IO_{11}^I)
\label{AJJJ}\\ &&
-\kappa(O^I_{31}O^I_{13}O^{I2}_{12}+O_{11}^IO_{12}^IO_{13}^IO_{32}^I)\Big]
AJ^3
\nonumber
\end{eqnarray}
which complete all the relevant information necessary to calculate
the production and invisible decay of the Higgs bosons.

\section{Numerical Expectations of the Model}

In this section we describe the expectations of our Model for the
various Higgs boson masses and couplings relevant for our discussion.
In order to do this we numerically diagonalize the mass matrix in
eq.~(\ref{MR}) and impose the minimisation conditions $t_i=0$, $i=1,2,3$,
where the tadpoles are in eq.~(\ref{eq:tadpoles}), 
and check the positivity of the three CP-even and
CP-odd eigenvalues. As seen explicitly, these matrices are determined
in terms of the 9 dimension-less coupling constants and the three VEVs
characterising the Higgs potential, as the three mass parameters
$\mu_i$ have all been eliminated. Note, however that there is a
restriction that arises from the $W$ mass constraint that relates
$v_2$ and $v_3$ through eq.~(\ref{mw}) that can be written as
\begin{equation}
\sqrt{v_2^2 + 2v_3^2} \simeq 246 \, \mbox{GeV}
\end{equation}
Moreover, $v_3$ must be smaller than about 8 GeV in order to obey
the experimental value of the $\rho$ parameter defined in eq.~(\ref{rho}).
 
\begin{figure}
\centerline{\protect\hbox{
\psfig{file=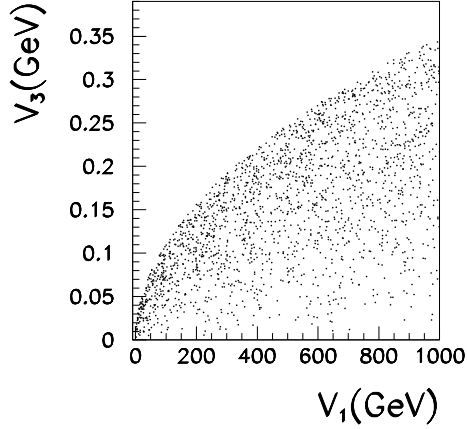,height=6truecm}}}
\caption{Allowed region in $v_1$--$v_3$ space obtained when the 
parameters are varied as described in the text.}
\label{v3-v1}
\end{figure}

A more stringent constraint on $v_3$ follows from astrophysics, due to
the stellar cooling argument, already mentioned. Indeed, if produced
in a stellar environment via the Compton-like process $\gamma + e
\to  J + e$, the majoron would escape leading to excessive
energy loss \cite{KIM}. In order to suppress this one must severely
restrict the coupling of the Majoron to the electrons. 
Such coupling arises from the projection of the majoron $J$
onto the doublet, leading to
\begin{equation}
|\langle J|\phi\rangle|=\frac{2|v_2|v_3^2}{
\sqrt{v_1^2(v_2^2+4v_3^2)^2+4v_2^2v_3^4+v_2^4v_3^2}} \lsim 10^{-6}
\label{AstroContraint}
\end{equation}
In order to have an idea of the parameter ranges involved we have
randomly varied over the parameters $0<\lambda_i<4$, $0<\beta_i<4$, (in
order to guarantee a perturbative regime) and the three $v_i$ subject
to the above restrictions, with the lepton number violating $v_1$
varied in the range $0 <v_1< 1000 \: \hbox{GeV}$.  The resulting
$v_1$-$v_3$ region allowed by the model is seen 
in Fig.~\ref{v3-v1}. The shape of the region can be understood easily from 
eq.~(\ref{AstroContraint}) noting that, since the lepton number violating 
VEV $v_3$ is small and $v_2$ is almost fixed by the $W$ mass constraint,
then the boundary of the allowed region satisfies $v_3\sim\sqrt{v_1}\,$.

\begin{figure}
\centerline{\protect\hbox{
\psfig{file=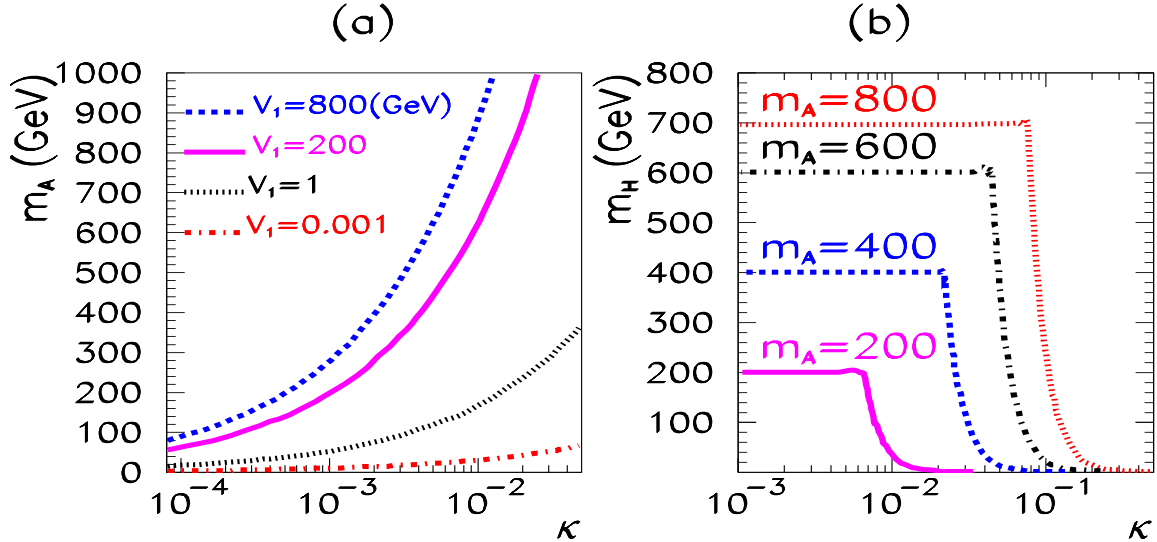,height=8.5truecm,width=16truecm}}}
\caption{Lowest-lying Higgs boson masses allowed in our model 
when the parameters are varied as described in the text.}
\label{zone2}
\end{figure}

The lowest-lying Higgs boson masses in our model may be similarly
determined after imposing the above restrictions.  For example, it is
instructive to display the results as a function of the parameter
$\kappa$ characterising the Higgs potential. The corresponding plots
are shown in Figs.~\ref{zone2}a--b. In Fig.~\ref{zone2}a we plot the
correlation between the massive pseudoscalar mass and the parameter
$\kappa$, for different values of $v_1$. Each curve corresponds to the
boundary of a scatter plot with the solutions concentrated above
it. Its shape can be understood from \eq{MA} and \eq{AstroContraint}
where we find $m_A\sim\sqrt{\kappa}$.  In order to have $m_A$ below a
certain value one requires an upper bound on $\kappa$, which tightens
with larger $v_1$. From Fig.~\ref{zone2}b, we can see that, as a
consequence of the smallness of $v_3$, $m_{H_1} < m_A$, except for a
narrow window in which $m_{H_1} \gsim m_A$. In such a small region the
decay $H_1\to AJ$ would be allowed, while the decay $H_1 \to AA$ is
forbidden by phase space. This can be contrasted with the MSSM where
the decay $h\to AA$ is allowed, though in a very small region in
parameter space \cite{haasusy}. When $\kappa$ is small $H_1$ is mostly
triplet and almost degenerate with $A$ and this corresponds to the
horizontal lines in Fig.~\ref{zone2}b. For larger $\kappa$ the
component of $H_1$ along the triplet decreases and $H_1$ becomes
lighter than $A$, as seen in Fig.~\ref{zone2}b.

\begin{figure}
\centerline{\protect\hbox{ 
\psfig{file=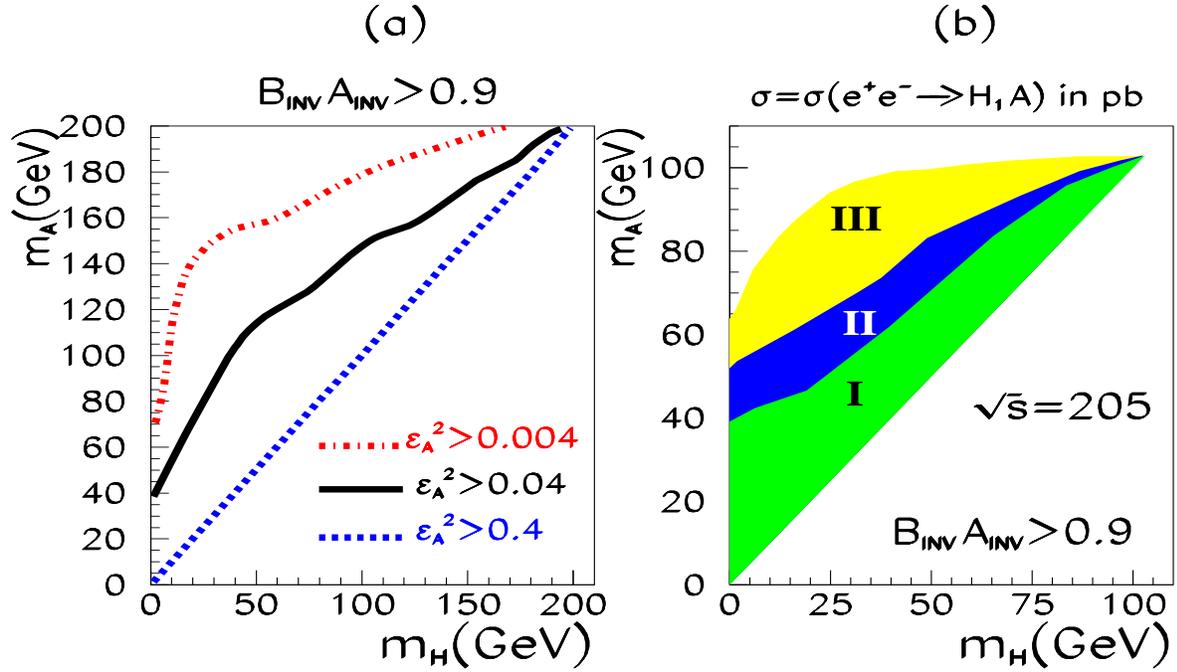,height=9truecm,width=16truecm}}}
\caption{Lowest-lying Higgs boson masses versus effective 
coupling strength parameter (a) and associated production 
cross section (b).}
\label{ma-mh-eps}
\end{figure}

\begin{figure}
\centerline{\protect\hbox{ 
\psfig{file=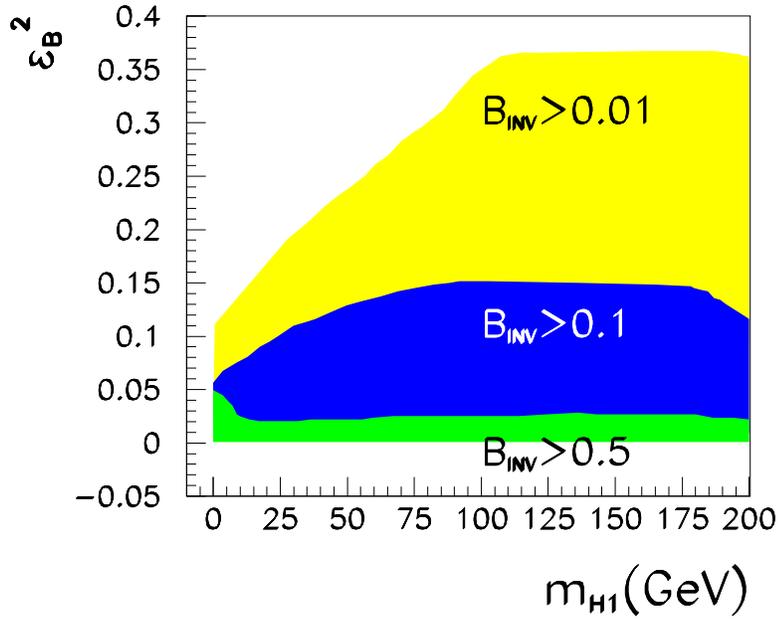,height=8.5truecm}}}
\caption{Bjorken production effective strength parameter
versus $m_{H_1}$ for different ranges of the parameter $B_{inv}$. }
\label{eps-mh}
\end{figure}

We have verified explicitly that in our model the invisible branching
ratios of $H_1$ and $A$ given by
\footnote{We have neglected the invisible decay $A \to \nu \nu$
relative to $A \to 3 J$, which is expected for reasonable choices
for the quartic parameters in the scalar Higgs potential and
for the lepton Yukawa couplings.}
\begin{eqnarray}
B_{inv}&=&BR(H_1 \to  JJ) + BR(H_1 \to  JA)
BR(A \to  JJJ),\nonumber\\
A_{inv}&=& BR(A \to  JJJ) + BR(A \to  JH_1)
BR(H_1 \to  JJ),
\label{hainv}
\end{eqnarray}
and their product $B_{inv}\,A_{inv}$, which will be needed in the next
section, can be large and even 100 \% over large regions of the
parameter space. This can be seen in Fig.~\ref{ma-mh-eps} where we are
considering only points in parameter space where
$B_{inv}\:A_{inv}>0.9$.  We also have verified that $\epsilon_A^2$ can
vary over all its range for all possible values of the invisible
branching ratios.  Thus, one may obtain plots similar to
Fig.~\ref{ma-mh-eps}a for other possible values of the product
$B_{inv}\,A_{inv}$.  The solutions where $\epsilon_A^2$ is larger than
the label associated to a particular curve are concentrated in the
region between the curve and the main diagonal. The points
corresponding to $\epsilon_A^2>0.4$ are so close to the main diagonal
that the width of the region cannot be seen with the naked eye. An
alternative way to display this information is in terms of the
associated production cross section which we choose to calculate at
$\sqrt s=205$ GeV. This is shown in Fig.~\ref{ma-mh-eps}b. In this
figure the diagonal line corresponds to the maximum cross section
$\sigma_{\rm max}\approx 0.5$ pb, while region I corresponds to points
where the cross section lies between 0.1 pb and 0.5 pb. For region II
we have $0.1>\sigma \geq 0.01$ pb and for region III, $0.01>\sigma
\geq 0.001$ pb. 
Note that there are no points with $m_{H_1}>m_A$ except very near the
line $m_A=m_{H_1}$.  For completeness we also present in
Fig.~\ref{eps-mh} the results for the Bjorken production. This plot
displays the effective coupling strength parameter $\epsilon_B^2$
versus $m_{H_1}$ for different ranges of the invisible branching ratio
$B_{inv}$.
Note that $B_{inv}$ is large only when the coupling of the lightest
CP-even Higgs to the fermions is small. This coupling is determined by
the projection of the lightest CP-even Higgs onto the doublet Higgs
boson, $O^R_{12}$. Should it be small the corresponding value of
$\epsilon_B\approx O^R_{12}$ which determines the Bjorken cross
section is also small. This correlation can easily be seen from
Fig.~\ref{eps-mh}. As a result if the Higgs is produced via the
Bjorken process it is likely to decay visibly, as in the SM.

\section{Model-independent Analysis}

In this section we perform a model independent study of the limits
that can be set based on Higgs boson production in $e^+e^-$ colliders
and its subsequent decays, focussing on LEP.  

Consider the massive pseudoscalar $A$ and the lightest CP--even scalar 
$H_1$. If $m_A>m_{H_1}$ then $A$ may decay in the standard way to 
$b\bar{b}$, or to $b \bar{b} + p\!\!\!/_T$ via $A \to H_1 J$ with $H_1
\to b \bar{b}$, or invisibly into three majorons. From unitarity it
follows that only two of these three branching ratios are
independent. In addition, the lightest scalar boson $H_1$ can decay
either to $b\bar{b}$ or invisibly to two majorons and only one of the
two branching ratios is independent.  Similarly, if $m_{H_1}>m_A$ we
have in total three independent branching ratios. Thus, in order to
make a model independent analysis, we need seven parameters to
describe the implications of the production of Higgs bosons at the $Z$
peak: two masses $m_A$ and $m_{H_1}$, the two parameters $\epsilon_A$
and $\epsilon_B$ which determine the Bjorken and associated production
cross sections and finally, three independent branching ratios (two
for the heavier Higgs boson and one for the lightest).  Table 1 shows
the signatures expected in the model for the cases of Bjorken and
Associated production. 

In order to have an idea one may compare the above seven parameters
with those needed in the simpler models considered before. In the
majoron-less model in ref.~\cite{TD} only five parameters would be
relevant, as there is a unitarity relation ${\epsilon_B}^2 +
{\epsilon_A}^2 =1$ which does not hold in the present case because the
admixture of the singlet Higgs bosons reduces the $H$ and $A$
couplings to the Z.  The present model has the additional $A \to H J$
branching ratio. Moreover in ref.~\cite{TD} the $A$ must decay either
visibly (mainly to $b\bar{b}$) or invisibly to neutrinos.  On the
other hand in the majoron model considered in ref. \cite{eboli} there
are also five parameters: $m_A$, $m_H$, $\epsilon_B$, $\epsilon_A$ and
finally, the visible $H$ decay branching ratio is an abitrary
parameter. Note that the $A$ must decay visibly (to $b\bar{b}$ mainly)
but there is no unitarity relation for the $\epsilon$'s due to the
admixture of the singlet Higgs bosons.

It is outside the scope of our present paper to perform an exhaustive
study of restrictions on the parameters of the Higgs potential of this
model, especially because of its complexity.  However we analyse all
signatures that can be engendered by Higgs boson production and its
subsequent decays in this model. Although tedious, it is a
straightforward task to convert the bare information we provide into
restrictions on the model parameters. However we prefer not perform
this in detail and use the underlying model only to motivate the
analysis.

First we study the constraints arising from the invisible $Z$ width,
following closely the analysis performed in ref.~\cite{TD}, where a
simpler model, with lepton number violation introduced explicitly and
the Higgs bosons decaying to neutrinos rather than to majorons was
analysed. As in the case of the model in ref.~\cite{TD}, the Bjorken
process contribution to the invisible $Z$ width $Z \to  Z^* H_1$
is very small. Therefore we consider the limits that can be set on
associated Higgs boson production at the Z peak, $e^+e^-
\to Z \to H_1 A$ when both CP-even ($H_1$) as well as CP-odd Higgs
bosons ($A$) decay invisibly. The contribution to the invisible $Z$
width in terms of invisible branching ratios $B_{inv}$ and $A_{inv}$
can be found in ref.~\cite{TD}, and the invisible branching ratios
themselves are defined in eq.~(\ref{hainv}).

\begin{figure}
\centerline{\protect\hbox{
\psfig{file=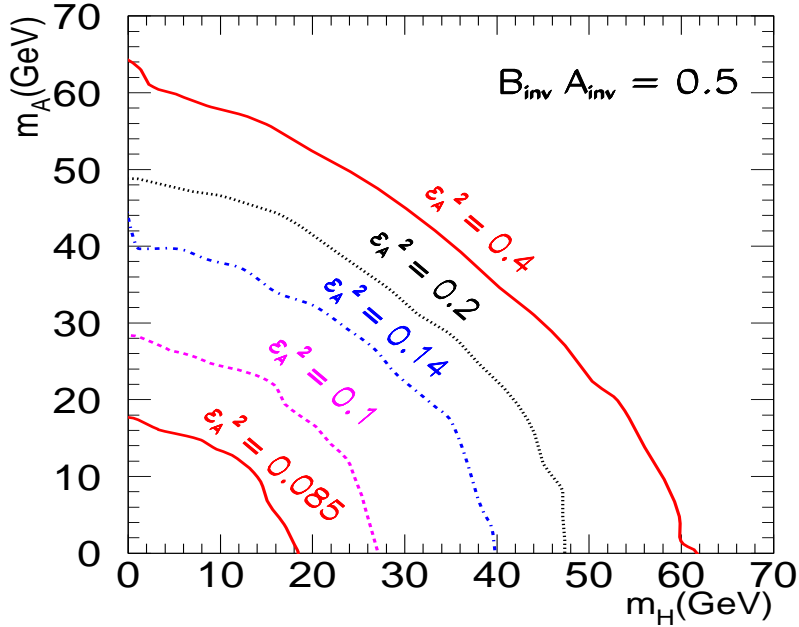,height=10truecm,width=12truecm}}}
\caption{95 \% CL bounds on ${\epsilon_A}^2$ in the $m_H$-$m_A$ plane for
the the indicated value of the invisible branching ratios}
\label{results2}
\end{figure}
In Fig.~\ref{results2} we show 95 \% CL bounds on ${\epsilon_A}^2$ in
the $m_{H_1}$-$m_A$ plane for a fixed illustrative value of the
product $B_{inv}A_{inv}=0.5$. Five curves labelled by a value of
$\epsilon_A^2$ are shown. No points below each of these curves are
allowed with $\epsilon_A^2$ larger than that value. The corresponding  
exclusion plot corresponding to $B_{inv}A_{inv}=1$ has been given in
ref.~\cite{TD}.  We see from these plots that simply by using the
neutrino counting at the Z peak one can already impose important
constraints on the parameters of the model.  For example, for $H_1$
and $A$ masses around 20 GeV the upper bound on ${\epsilon_A}^2$ is a
few times $10^{-2}$. 

Apart from an additional contribution to the invisible $Z$ width, the
model produces the variety of signals shown in table 1. Most of these
are exactly the same as analysed in \cite{eboli}. Though the analysis
presented in ref. \cite{eboli} was in a different context, those
results are applicable here. They are summarized in Figs 4 and 5 of
ref.~\cite{eboli}. These plots may be regarded as particular cases of
our model when $A_{inv} \to 0$. With appropriate re-interpretation
they can be adapted to our case. However, as we mentioned, we will not
enter into that.

We now consider the various final state topologies that can be
produced in $e^+e^-$ collisions at LEP, for example those exhibiting
$b \bar{b}$ or $\ell^+ \ell^-$ ($\ell=\mu$ or $e$) pairs and missing
energy. The complete table of signatures is reproduced in table 1.

\begin{table}[t]
\label{signals}
\begin{center}
\begin{tabular}{cccc}
 &Associated production & Bjorken production & \\ \hline
 &$b \bar{b} b \bar{b} $ & $b \bar{b} p\!\!\!/_T $&  \\
 &$b \bar{b} p\!\!\!/_T$ & $b \bar{b} l^+l^- $& \\
 &$b \bar{b} b \bar{b}p\!\!\!/_T$ & $b \bar{b} q \bar{q}$& \\
 &$p\!\!\!/_T$ & $l^+l^- p\!\!\!/_T$& \\
& &$l^+l^- p\!\!\!/_T$& \\
&  &$p\!\!\!/_T$ &\\
\end{tabular}
\end{center}
\caption{Final signals arising from associated
production (left column) as well Bjorken production (right column).}
\end{table}

A lot of information follows from the detailed study of these signals.
Of the topologies considered in table 1, all have been previously
analysed in ref. \cite{eboli} by carefully evaluating the signals and
backgrounds, and by choosing appropriate cuts to enhance the discovery
limits. There is only one exception: the present model contains a
completely novel signature, namely four b-jets plus missing
momentum. This is not present in \cite{eboli} nor in \cite{TD}. As far
as we know it is the first extension of the Higgs sector with this
feature. Therefore, from now on we concentrate on this $b \bar{b}
b\bar{b}p\!\!\!/_T$ signal. The main background comes from $e^+e^-
\to Z\gamma Z\gamma $, $e^+e^- \to WW$ and $e^+e^- \to Z\gamma $.
This background has been analysed in other contexts, such as chargino
production at LEPII \cite{navas,tauj}, where two charginos are
produced decaying each one into a neutralino plus a $W$ boson, where
the neutralino is stable or decays invisibly.  This also gives the 4
jets + missing momentum signature. With appropriate cuts in the
$p\!\!\!/_T$, number of jets and invariant mass distributions the
background could be removed keeping high signal efficiencies.  For our
illustrative purposes, we imposed the following cuts in order to
remove the background:
\begin{itemize}
\item 
when dealing with hadronic events we only select those events with at
least 12 charged particles in the final state.  
\item 
In to avoid high energy initial state radiation $Z$ events we reject 
events with a photon with an energy of more than 35 GeV
\item 
We only accept events with at least four jets.
\item 
We reject an event if the sum of the energy of the two less energetic
jets is less than 10 GeV, in order to remove events like $e+e-
\to q{\bar q}g$, and $e+e- \to (Z\to q {\bar q})(\gamma* \to q{\bar q})$
characterised by two very energetic jets plus two less energetic ones.
\item 
We reject events with a missing transverse momentum smaller than
10 GeV, in order to avoid events with particles going down the beam
pipe.  
\item 
We finally impose the invariant mass of the event to $m_{inv} > 100$
GeV.  This cut is essential to kill the $Z$ background, which has a
large cross section.
    
\end{itemize}
  
\begin{figure}
\centerline{\protect\hbox{
\psfig{file=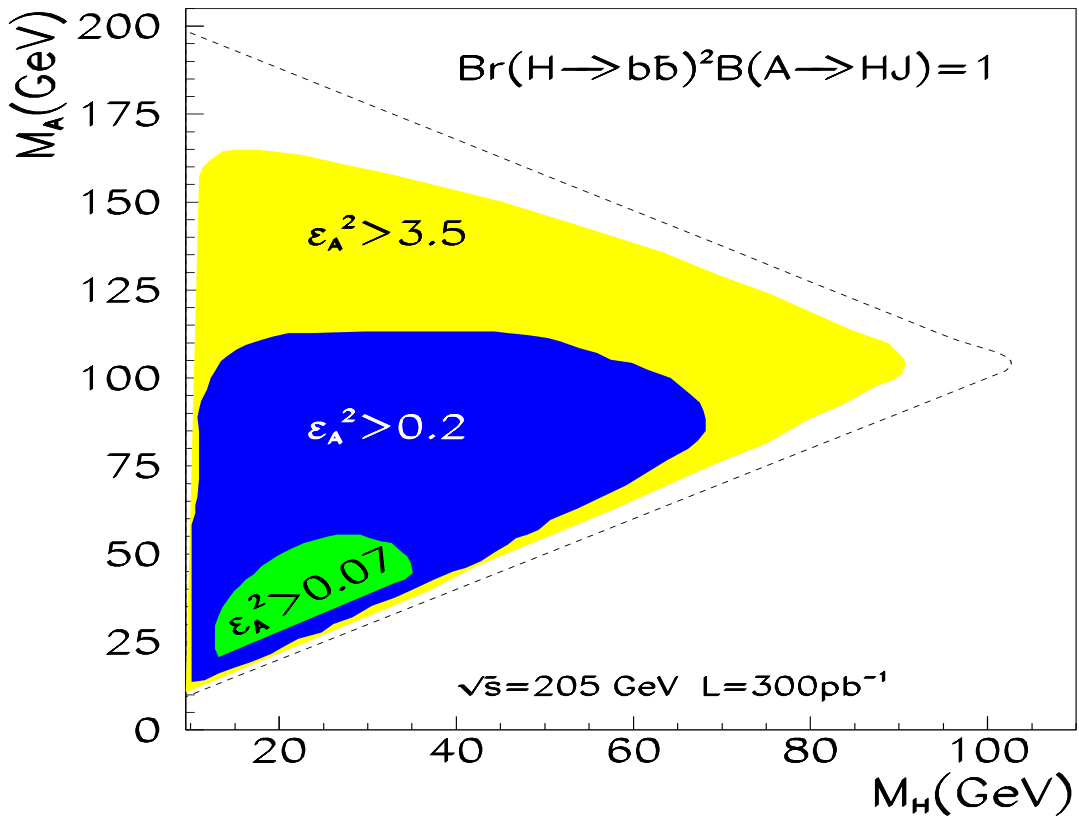,height=12truecm,width=16truecm}}}
\caption{95 \% CL bounds on ${\epsilon_A}^2$ in the $m_H$-$m_A$ 
plane that follow from the 4jets + missing transverse momentum
analysis.}
\label{4bptmiss}
\end{figure}

Applying all these cuts we eliminate the background, that has been
simulated using JETSET \cite{pythia}. We did a Monte-Carlo study of
the signal for our model, which allowed us to calculate the efficiency
for the signature after implementing the above mentioned cuts.  Our
results are given as a 95 \% CL exclusion plot in the $M_H-M_A$
plane shown in Fig.~\ref{4bptmiss}. We have assumed a LEPII integrated
luminosity of 300 $pb^{-1}$.
These results are complementary to those arising from the invisible width
only, and also complement those that can be derived by appropriate
rescaling of the plots shown in \cite{eboli} corresponding to the other
signals in table 1.

\section{Conclusions}

We have studied the neutral Higgs sector of the general seesaw
majoron--type model of neutrino mass generation with spontaneous
violation of lepton number. This sector contains three massive CP--even
Higgs bosons $H_i$, $i=1,2,3$, one massive CP--odd $A$, and one
massless CP--odd $J$ called the majoron. We show that $H_1$ and $A$
can decay invisibly into majorons and determine the constraints that
arise from the invisible decay width of the $Z$ gauge boson.  The
existence of such novel invisible pseudoscalar Higgs boson decays
discussed in this paper should be taken into account when determining
the Higgs boson discovery prospects at LEP II and other colliders,
such as the LHC and NLC. 

We have also noted that the existence of the new decay channel $A \to
H_1 + J$ leads to a novel four-jet + missing momentum signature in
associated Higgs boson production $e^+e^- \to H_1 A$ when $A
\to H_1 J$ and $H_1 \to b \bar{b}$. This could be detectable at LEP
II.  We have studied the background and performed a Monte-Carlo
simulation of the signal in order to determine the limits that follow
from that. Although the structure of the Higgs sector is quite rich
one has already important restrictions on Higgs boson mass, couplings
and branching ratios that should be taken into account in relation to
possible new studies at future colliders such as such as the LHC
\cite{lhc} and the NLC \cite{valle:2} or at a possible muon collider.

\section*{Acknowledgements}

We thank A. Joshipura, J. J. Hernandez and S. Navas for useful
discussions.  Special thanks to Oscar Eboli for very helpful
discussions related to the study presented in section VI. This work
was supported by DGICYT under grant PB95-1077 and by the TMR network
grant ERBFMRXCT960090 of the European Union. M. A. D. was supported by
a DGICYT postdoctoral grant, D. A. R. was supported by Colombian
COLCIENCIAS fellowship, while M. A. G-J was supported by a Spanish MEC
FPI fellowship.

\end{document}